\documentclass[aps,prx,twocolumn,nofootinbib,superscriptaddress,showpacs,showkeys,longbibliography]{revtex4-2}

\usepackage{amsmath, amsxtra, amssymb, mathtools, isomath, nccmath, xspace, siunitx,graphicx}
\usepackage{color,hyperref,sidecap,natbib,dsfont,xcolor,pifont}
\usepackage[utf8]{inputenc}
\usepackage[british]{babel}

\newcommand{\ket}[1]{\ensuremath{\lvert #1 \rangle}\xspace}%

\sidecaptionvpos{figure}{h}
\long\def\symbolfootnote[#1]#2{\begingroup%
\def\thefootnote{\fnsymbol{footnote}}\footnotetext[#1]{#2}\endgroup}

\makeatletter
\@addtoreset{equation}{myequation}
\makeatother

\begin{document}
\setlength{\belowdisplayskip}{9pt}
\setlength{\abovedisplayskip}{9pt}
\setlength{\belowdisplayskip}{9pt}
\setlength{\abovedisplayskip}{9pt}

\title{\bf{Rydberg Macrodimers: Diatomic molecules on the micrometer scale}}

\author{Simon Hollerith}
\email[]{Shollerith@fas.harvard.edu}
\affiliation{Max-Planck-Institut f\"{u}r Quantenoptik, 85748 Garching, Germany}
\affiliation{Munich Center for Quantum Science and Technology (MCQST), 80799 Munich, Germany}

\author{Johannes Zeiher}
\affiliation{Max-Planck-Institut f\"{u}r Quantenoptik, 85748 Garching, Germany}
\affiliation{Munich Center for Quantum Science and Technology (MCQST), 80799 Munich, Germany}
\date{\today}

\begin{abstract}
Controlling molecular binding at the level of single atoms is one of the holy grails of quantum chemistry.
Rydberg macrodimers -- bound states between highly excited Rydberg atoms -- provide a novel perspective in this direction.
Resulting from binding potentials formed by the strong, long-range interactions of Rydberg states, Rydberg macrodimers feature bond lengths in the micrometer regime, exceeding those of conventional molecules by orders of magnitude.
Using single-atom control in quantum gas microscopes, the unique properties of these exotic states can be studied with unprecedented control, including the response to magnetic fields or the polarization of light in their photoassociation. The high accuracy achieved in spectroscopic studies of macrodimers makes them an ideal testbed to benchmark Rydberg interactions, with direct relevance to quantum computing and information protocols where these are employed. 
This review provides a historic overview and summarizes the recent findings in the field of Rydberg macrodimers. 
Furthermore, it presents new data on interactions between macrodimers, leading to a phenomenon analogous to Rydberg blockade at the level of molecules, opening the path towards studying many-body systems of ultralong-range Rydberg molecules.


\end{abstract}
\maketitle

\section{Introduction}
Strong interactions between Rydberg atoms enabled numerous groundbreaking experiments in quantum sciences and technologies.
Rydberg macrodimers~\cite{Boisseau2002,Overstreet2009,Sassmannshausen2016,Hollerith_2019} provide the most precise platform to study these interactions and enable observations of basic properties of molecules at an exceptional level of control.
Experimental studies of molecules are challenging due to their small size and typically random orientation within experimental samples~\cite{reaction_gas_mic,Stapelfeldt_alignment,Denschlag_alignment,Kunitski2021}.
Furthermore, \emph{ab initio} calculations are difficult and require sophisticated computational methods, even for diatomic molecules~\cite{Quantum_diatomics_1,Quantum_diatomics_2,Quantum_diatomics_3,Quantum_diatomics_4,CaF_AbInitio} and in particular, if they consist of many-electron atoms.
Experiments at ultracold temperatures enabled studies of molecules at a new level of precision~\cite{chem_rev_coldmols_2016,bohn_cold_2017,hu_direct_2019,rovib_ground_first,Zelevinsky16}.
The high level of control and the low energy scales in cold atom systems also provided the experimental ground for the observation of Rydberg macrodimers~\cite{Overstreet2009,Sassmannshausen2016}.
The large bond lengths and the small binding energies of macrodimers enable microscopic access to the atoms forming the molecule and studying and shaping their electronic structure~\cite{Hollerith_2021}.
Because mainly two highly-excited electrons~\cite{Weber2017}, separated well from the remaining atomic constituents, are involved in the binding, their theoretical description inherits the simplicity of Rydberg atoms and their vibrational and electronic structure can be calculated at high precision.
In reverse, resolving their vibrational structure with high resolution reveals new details about Rydberg atoms and their interactions, such as the presence of hyperfine interactions in Rydberg pair potentials~\cite{Hollerith_2021} or non-adiabatic motional transitions between different Born-Oppenheimer potentials~\cite{Hollerith_2019}. 

\begin{figure*}[htp]
  \centering
  \includegraphics[width=1.65\columnwidth]{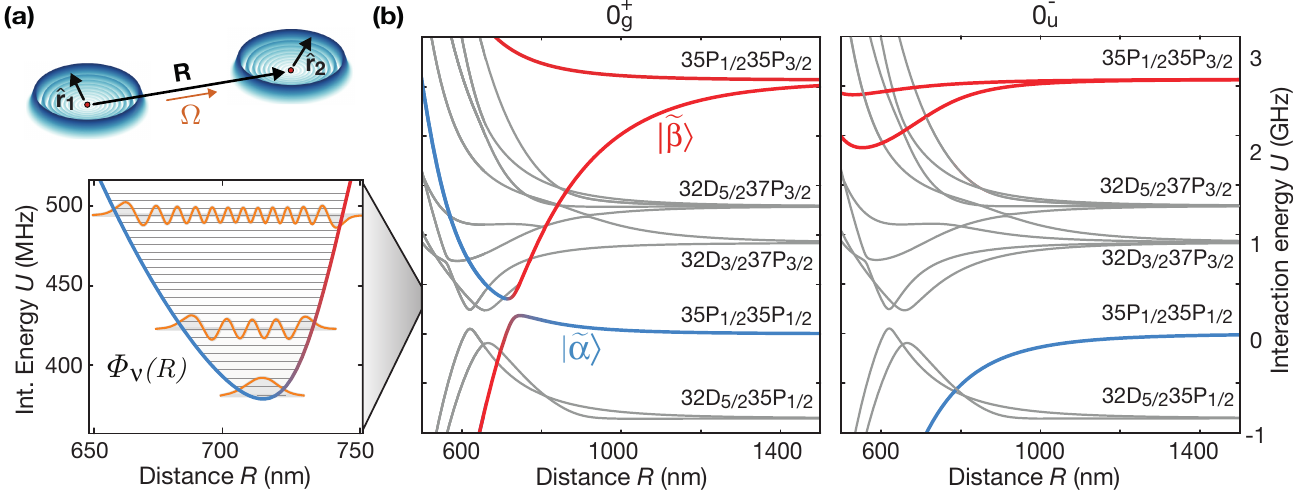}%
  \caption{\label{fig:schematic}
  \textbf{Overview over Rydberg macrodimers.} Figure partially adapted from~\cite{Hollerith_2019}. \textbf{(a)} Macrodimers are electrostatically bound highly-excited Rydberg atom pairs. The binding is mediated by the exceptionally large dipole moments of Rydberg atoms and occurs at interatomic distances $R = |\mathbf{R}|$ where the orbitals of both Rydberg electrons (blue), located at distances $\hat{\mathbf{r}_1}$ and $\hat{\mathbf{r}}_2$ from both ionic cores (red dots), do not overlap.  Their micrometer-large bond length makes them the largest existing electrically neutral diatomic molecules. \textbf{(b)} The binding potentials typically form because of avoided crossings between Rydberg pair potentials. In the shown example for $^{87}$Rb pairs excited to Rydberg P-states, the two crossing potentials $\ket{\widetilde\alpha}$ and $\ket{\widetilde\beta}$ are coupled by the dipole-dipole interaction Hamiltonian $\hat{H}_{\mathrm{int}}(R)$. The blue~(red) color indicates the $R$-dependent adxmixture of the pair state $\ket{\widetilde\alpha}$~($\ket{\widetilde\beta}$) of the potential. In the asymptotic limit of large distances, the interactions are typically of van der Waals type. Due to the symmetry of the interaction Hamiltonian, the pair potentials decouple into different branches $|\Omega|^\pm_{g/u}$ which are labelled by their angular momentum projection $\Omega$ on the interatomic axis as well as their reflection (superindex) and inversion symmetry (subindex).}
\end{figure*}

\subsection{Classification}
Rydberg macrodimers belong to the class of \emph{purely long-range molecules} (PLRM) which have been first predicted in the 1970's~\cite{PLR_theory}. 
These are molecules where the overlap of electronic orbitals vanishes over the full extension of the binding potential~\cite{Julliene_Review06}. 
This differs from conventional deeply bound molecules where electrons occupy hybridized orbitals delocalized over the whole system~\cite{Demtroeder_rotation_2007}. 
Even for weakly bound complexes of noble gases $\mathrm{He}_2$~\cite{Helium_dimer_1,Helium_dimer_2,blaney_van_1976} where such chemical bonds do not form, the orbital overlap becomes relevant for the repulsive potential barrier at short distances~\cite{vdW_Dimer_rep_barrier,vdW_Dimer_rep_barrier2}.
PLRM typically form at avoided crossings between different asymptotic pair states in the large-distance limit~\cite{PLR_theory_0}.
They have been first observed at lower principal quantum numbers in the 1990's~\cite{PLR_exphf,pure_long_range_1,pure_long_range_2,PLR_helium}. 
As for macrodimers, binding potentials can be calculated with high accuracy~\cite{PLR_exphf}.
This enabled precision tests between theory and experiment and led to the observation of retardation on the binding potentials~\cite{Retardation_PLRM_1996}. 

Macrodimers furthermore can be further classified as \emph{ultralong-range Rydberg molecules} (ULRM)\,\cite{Shaffer2018}.
The class summarizes molecules where the presence of Rydberg atoms leads to bond lengths orders of magnitude larger than usually.
In addition to macrodimers, the class contains bound states between ground state atoms and a single Rydberg atom~\cite{Ultra_long_range_theory,Bendkowsky2009,fey_ultralong-range_2020,SR_Rydberg_molecule_polaron}.
The field of ULRM has recently been extended by bound states between a Rydberg atom and an ion~\cite{ion_Ryd_mol_2022}.

\subsection{Properties of macrodimers}
The micrometer-sized bond lengths of macrodimers, as large as small bacteria or the wavelength of visible light, are the largest among all types of electrically neutral diatomic molecules.
Compared to most other molecules, the absence of overlapping electron orbitals significantly simplifies calculations of the pair potentials \emph{ab initio} from atomic properties~\cite{Hollerith_2019,Hollerith_2021}. 
Besides the absence of explicit electron correlations, macrodimers have key signatures of molecules, such as vibrational and rotational degrees of freedom or conserved quantities of the electronic state related to the symmetries of their point group~\cite{Demtroeder_rotation_2007}. 

Their large size originates from the large dipole moment of Rydberg atoms, which can easily reach the kilodebye regime, as well as the small energy separation between neighboring Rydberg states~\cite{gallagher_rydberg}.
Furthermore, the large polarizability of Rydberg states provides large tunability using external fields.
The depth of the binding potentials is typically a few hundred megahertz, the vibrational splittings are on the order of a few megahertz~\cite{Samboy2011}. 
Close to the potential minimum, the binding potentials can be usually well approximated by a harmonic oscillator potential.
The total number of vibrational modes is similar to conventional deeply bound molecules close to the electronic ground state. 
The lifetime of macrodimers is fundamentally limited by the radiative decay of the constituent Rydberg atoms.
Because the timescale of a molecular rotation typically exceeds the lifetime, macrodimers keep their spatial orientation until they decay.

The symmetries of macrodimers are the same as for any diatomic molecule~\cite{lefebvre-brion_chapter_2004}.
For homonuclear macrodimer states in the absence of external fields, the relevant point group is $D_{\infty h}$.
Macrodimers are best described by Hund's case~(c), where the molecular states are labelled by $|\Omega^{\pm}_{g/u}|$~\cite{Stanojevic2006,Theory_symmetries,Weber2017,Sibalic2017}. 
The total electronic angular momentum projection $\Omega = \Lambda + \Sigma$ on the interatomic axis $\mathbf{R}$ is conserved because of the symmetry of the two-atom system. 
The superscript (subscript) specifies the reflection (inversion) symmetry of the molecular state.
In contrast to many deeply bound molecules, falling in other Hund's cases where the binding depends critically on the orbital angular momentum projection $\Lambda$, the total spin projection $\Sigma$ and also $\Lambda$ are not conserved. 
Furthermore, the rotational energies of macrodimers are negligibly small and the rotational angular momentum remains uncoupled from the other contributing angular momenta.

\section{Theoretical description}\label{sec:theoretical_Des}
Macrodimers were theoretically predicted in 2002\,\cite{Boisseau2002}. 
Their binding potentials are calculated at interatomic distances $R \gg R_{\mathrm{LR}}$ larger than the so-called Le Roy radius $R_{\mathrm{LR}}$ of the atom pair where the spatial overlap of both electron orbitals vanishes~\cite{le_roy_long-range_1974}.
The binding occurs at distances about ten times larger than the extension of the Rydberg orbits.
The absence of electron exchange greatly simplifies the calculations. 
When two Rydberg atoms approach each other, the Hamiltonian consisting of the two individual atoms gets perturbed by their interaction. 
The interaction Hamiltonian between the two Rydberg atoms
\begin{align}
\hat{H}_{\textrm{int}}(\mathbf{R}) =  & \frac{e^2}{4\pi\epsilon_0} \Bigg( \frac{1}{|\mathbf{R} + \hat{\mathbf{r}}_2 - \hat{\mathbf{r}}_1|} + \frac{1}{|\mathbf{R}|} \nonumber \\ & - \frac{1}{|\mathbf{R}  - \hat{\mathbf{r}}_1|} - \frac{1}{|\mathbf{R}+\hat{\mathbf{r}}_2|} \Bigg) 
\end{align}
contains four Coulomb interaction terms~\cite{Weber2017}, see Fig.~\ref{fig:schematic}.
These represent the attraction between the first (second) Rydberg electron and the positively charged second (first) ionic core and the repulsion between both Rydberg electrons and both ionic cores.
At large distances where orbitals do not overlap, $\hat{H}_{\mathrm{int}}(\mathbf{R})$ can be efficiently expressed in a multipole expansion~\cite{singer_long-range_2005,CS_calculations_2006,Deiglmayr2016,flannery_long-range_2005,Weber2017}
\begin{equation}\label{eq:multipole_expansion}
\hat{H}_{\textrm{int}}(\mathbf{R}) = \sum^\infty_{\kappa_1,\kappa_2 = 1}\frac{\hat{H}_{\kappa_1,\kappa_2}}{4\pi\epsilon_0 R^{\kappa_1+\kappa_2+1}}.
\end{equation}
The different multipole terms $\hat{H}_{\kappa_1\kappa_2} \propto \hat{r}^{\kappa_1}_1\hat{r}^{\kappa_2}_2$ depend on the radial coordinate $\hat{r}_{1/2}$ as well as spherical harmonics accounting for the angular coordinates of the two individual Rydberg atoms. 
Because Rydberg wave functions $\ket{r_i}$ can be calculated with high accuracy using quantum defect theory, also the matrix elements of the multipole expansion terms can be evaluated with high precision.
Truncating the sum in Eq.~\ref{eq:multipole_expansion} after accounting for sufficiently many terms and diagonalizing the Hamiltonian for varying distances $R$ provides the relevant Born-Oppenheimer potentials $V(R)$.
For alkali atoms, precise calculations can be performed using available open-source software~\cite{Weber2017,Sibalic2017}.

At large distances, the potentials are typically described by van der Waals potentials $V(R) = C_6/R^6$~\cite{Deiglmayr2016}.
In this regime, interactions arise from the lowest-order multipole term $\hat{H}^{(3)}_{\textrm{int}}(\mathbf{R}) \propto R^{-3}$ of Eq.~\ref{eq:multipole_expansion} in second-order perturbation theory.
Because van der Waals coefficients of Rydberg atoms are extraordinarily large and Rydberg states are energetically close, different van der Waals potentials eventually cross at shorter distances.
In the presence of a finite coupling between the crossing potentials, the upper part of the avoided crossing realizes a binding potential if the gap is dominating over the vibrational energy scale. 
The electronic structure of macrodimers can be expressed by decomposing the electronic wave function corresponding the binding potential into non-interacting Rydberg pair states $\ket{r_i r_j}$ via~\cite{Hollerith_2021}
\begin{equation}\label{eq:pair_pot_exp}
|\Psi_{\textrm{el}}(R) \rangle = \sum_{ij} c_{ij}(R)|r_i r_j \rangle.
\end{equation}
The vibrational energies are obtained by calculating the motional eigenenergies within the binding potential. 
Macrodimer states can be expressed as
\begin{equation}\label{eq:pair_pots}
|\Psi^\nu_{\textrm{Mol}}\rangle = \Phi_\nu(R) |\Psi_{\textrm{el}}(R) \rangle,
\end{equation}
with $\Phi_\nu(R)$ the vibrational states~\cite{Hollerith_2019}.
As discussed later, the vibrational energies as well as the electronic structure can be experimentally probed.

In the presence of electric or magnetic fields, the coupling terms to the field are added to Eq.~\ref{eq:multipole_expansion} before diagonalization. 
For field components perpendicular to $\mathbf{R}$, where the rotational symmetry is broken and $\Omega$ not conserved, the number of required basis states is significantly larger. 

Previous studies and also this review covers homonuclear macrodimers of alkali atoms.
Recently, there is increasing interest in Rydberg states of atoms with more than one electron in the outer orbital, such as Sr~\cite{Sr_Rydspectroscopy,Endres_2020}, Yb~\cite{Yb_Ryd_spec,Yb_qubit}, or Er~\cite{Trautman_ErbSpec}.
Here, Rydberg transitions depend on the coupled spin state between the Rydberg-excited electron and the electrons in lower orbitals.
One finds different Rydberg series for the different multiplets, whose finite coupling can be described by multi-channel quantum defect theory. 
This leads to a higher density of pair potentials with richer substructure, in particular in the presence of hyperfine interactions such as for $^{87}\mathrm{Sr}$~\cite{Strontium_87_2020_calc} or external fields where this coupling becomes larger~\cite{Vaillant_2012}. 
For these cases, vibrationally resolved studies of macrodimers will be helpful to test recently developed theoretical frameworks~\cite{Vaillant_2014,Strontium_87_2020_calc}.

\subsection{Scaling laws}\label{properties}
Many properties of Rydberg atoms can be calculated using their characteristic dependence on the effective principal quantum number $n^\star = n - \delta(n,L,J)$~\cite{gallagher_rydberg}.
Here, $n$ is the principal quantum number of the Rydberg state, $L$ ($J$) the orbital (total electronic) angular momentum. Furthermore, $\delta(n,L,J)$ are the mainly $L-$dependent quantum defects which are obtained from Rydberg spectroscopy~\cite{gallagher_rydberg,Sibalic2017}. 

Similar approximate scaling laws exist for macrodimers~\cite{Samboy2011,Barbier2021}.
Precise calculations of macrodimers again require detailed knowledge of $\delta(n,L,J)$ because they determine the properties of the contributing Rydberg levels.
The binding energies of macrodimers and their vibrational frequencies typically scale as $U_b \propto (n^\star)^{-3}$. 
For increasing $n$, the bond length $R_b \propto (n^{\star})^{8/3}$ increases faster than the separation of the Rydberg electron from the ionic core which scales as $\propto (n^{\star})^{2}$.
Assuming a rigid rotor, the rotational energy is given by $E_r = h B_r \ell (\ell + 1)$, with $\ell$ the rotational quantum number. 
 Because of the large bond lengths, the rotational constant $B_r = \hbar^2/(2\mu R_b^2)$ is typically below a kilohertz~\cite{Hollerith_2021}, lower than the expected decay rate.
Because macrodimers are bound at a distance where electrons do not overlap, auto-ionization rates are small.
In many cases, their lifetime is limited by the lifetime of the contributing Rydberg states~\cite{Boisseau2002,schwettmann_analysis_2007}, typically a few tens of microseconds.
In addition to a radiative decay $\gamma_{\mathrm{dec}} \propto (n^\star)^{-3}$ to the ground state, Rydberg atoms can also be transferred to other nearby Rydberg states by absorbing thermal photons.
These black-body transition rates decrease as  $\gamma_{\mathrm{bb}} \propto (n^\star)^{2}$.

In the presence of non-adiabatic motional couplings between different Born-Oppenheimer potentials, the macrodimer lifetime can be shorter, see also section~\ref{magn_predis}.
In particular for couplings to attractive potentials, the Rydberg pairs might eventually reach distances below the Le Roy radius where auto-ionization occurs~\cite{Ionization_Gallagher,Amthor_mechanical_ionization,hahn_ionization_2000,robicheaux_ionization_2005}.

\section{First observations}
Because of their small binding energies and short lifetimes, macrodimers can only be studied in isolated environments such as in cold atomic samples prepared in vacuum chambers and manipulated using laser beams.
A first step towards the observation of macrodimers is to measure the presence of interactions between Rydberg atoms. 
Rydberg interactions were first observed in the late 1980s as a broadening mechanism in Rydberg spectroscopy~\cite{J_M_Raimond_1981,Resonant_collision_spec_gal,Gallagher_dipolar_ex_1,Mourachko1998,Tong2004,asymmetric_broad_weidemueller}.
A second step is to probe the interactions in the non-perturbative regime at shorter distances, larger interaction shifts, and interaction-induced mixing of the electronic structure.
In 2003, researchers observed spectroscopic signatures several gigahertz detuned from the single-photon UV transition from the ground state $\ket{g} = \ket{5S_{1/2}}$ to Rydberg states $ \ket{nP_{3/2}}$ for $^{85}\,$Rb atoms trapped in a magneto-optical trap (MOT)~\cite{Macrodimers_1,Stanojevic2006,Theory_symmetries}.
The frequency agreed with the energy of asymptotic pair states $\ket{(n-1)D,nS}$ which were inaccessible by the UV laser due to dipolar selection rules.
However, dipole-dipole interactions admixed accessible Rydberg pair states at the avoided crossing point with a second van der Waals potential asymptotically connected to $\ket{nP_{3/2},nP_{3/2}}$~\cite{Optical_coupling_macrodimers}.
This observation agrees with more recent studies~\cite{Sassmannshausen2016} where excitation rates into pair potentials at avoided crossings are generally enhanced.
Here, because the pair potentials reach a local extremum, the motional-state overlap between the initial state and the excited Rydberg pairs is larger. 
Similar studies at higher spectral resolution in $^{133}\,$Cs also observed avoided crossings originating from weaker dipole-quadrupole interactions~\cite{dipole_quadrupole_dmayer2014}.
While these experiments demonstrated the presence of Rydberg interactions, they did not prove the presence of molecular-bound states. 
\subsection{Kinetic energy of ionized Rydberg atoms}
\begin{figure}
  \centering
  \includegraphics[width=0.92\columnwidth]{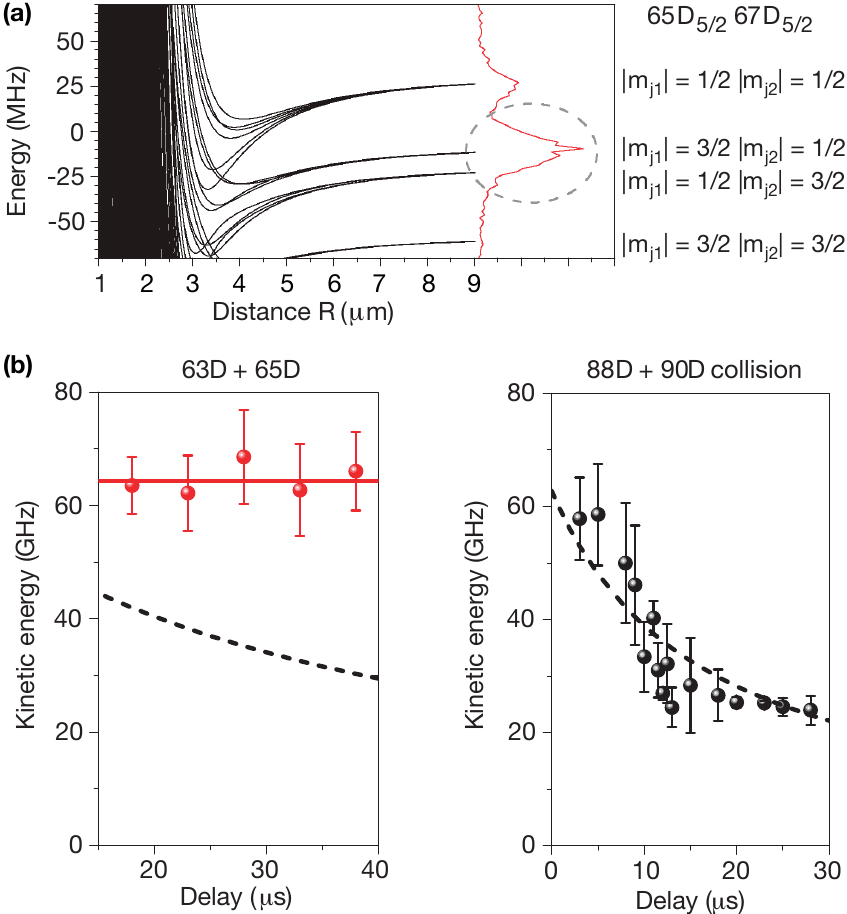}%
  \caption{\label{fig:Shaffer}
  \textbf{Observation via time-dependent distance distribution.} \textbf{(a)} Ground state atom pairs were laser-excited into Rydberg pair states $|(n-1)D_J,(n+1)D_J\rangle$ and detected as ions after ionization. \textbf{(b)} For pair states where calculations indicated the presence of macrodimer potentials, the kinetic energy of the ions was independent of the waiting time between Rydberg excitation and ionization (red) -- as expected for bound objects. In the absence of a binding potential, the energy of the ions decreases because the interactions of the Rydberg pairs before ionization increases the distance of the created ions (black). Figure adapted from~\cite{Overstreet2009}. }
\end{figure}

The first observation of macrodimers was reported in 2009~\cite{Overstreet2009} for $^{133}$Cs by studying two-photon resonances between excited pairs $\ket{6P_{3/2},6P_{3/2}}$ and Rydberg pair states $\ket{(n-1)D_J,(n+1)D_J)}$ in a  MOT~\cite{schwettmann_analysis_2007}, see Fig.~\ref{fig:Shaffer}. 
The atoms were excited into the short-lived states $\ket{6P_{3/2}}$ by the MOT light.
The Rydberg pair states were energetically isolated and well separated from other single-atom Rydberg states.
Pair potential calculations close to the asymptotic pair-state energies showed the presence of stable binding potentials.
The applied electric field strength~\cite{CS_calculations_2006,Stanojevic2006,cabral_effects_2011} and the principal quantum number $n \approx 65$ were chosen such that the potentials were accessible from the initial distance distribution.
After excitation, the Rydberg atoms were ionized using pulsed-field ionization (PFI).
The kinetic energy $E^k_\mathrm{ion}$ of the detected ions carries information about their distance before ionization due to their electrostatic repulsion. 
At the pair-state resonance, $E^k_\mathrm{ion}$ was independent of the waiting time $t_w$ between the excitation and ionization, showing the absence of forces that affect the interatomic distance.
The data instead indicated the presence of a force that stabilizes the interatomic distance - such as for a binding potential.
In contrast, tuning the laser to a dissociating pair state, $E^k_\mathrm{ion}$ decreased with $t_w$\cite{Overstreet2009,photoinitited_col_2007}.
The temporal spread of the ion signal from where $E^k_\mathrm{ion}$ was extracted also showed first indications of an alignment of the excited Rydberg pairs. 
\begin{figure}
  \centering
  \includegraphics[width=0.999\columnwidth]{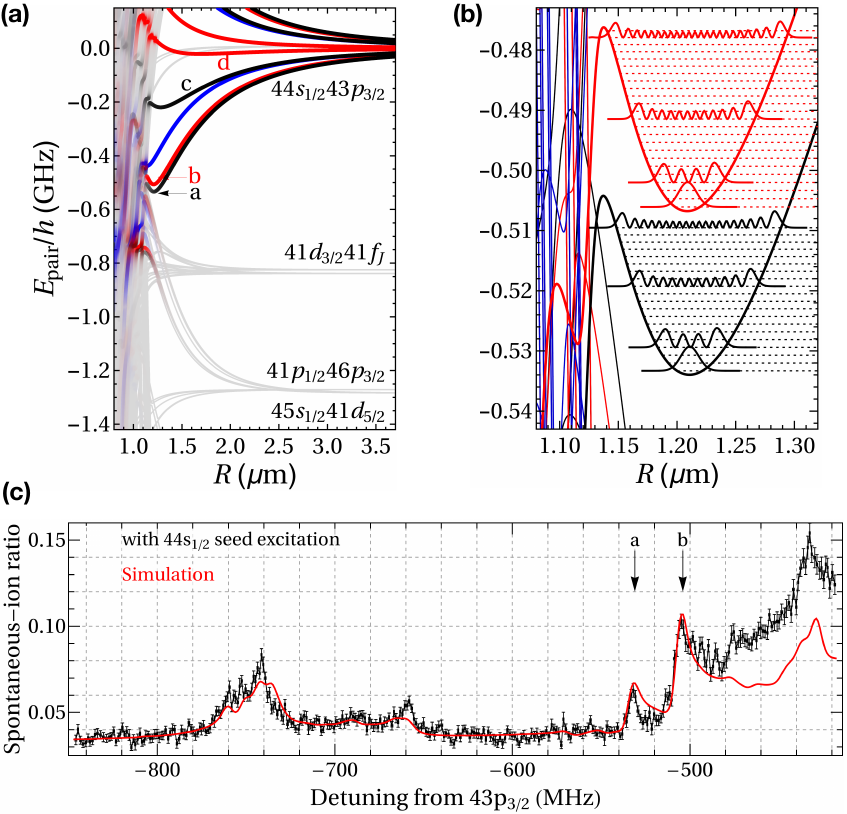}%
  \caption{\label{fig:Deiglmayr}
  \textbf{Spectroscopic identification of individual binding potentials.} \textbf{(a)} Calculations predict binding potentials close to the pair-state energy of the asymptotic state $\ket{44S_{1/2},43P_{3/2}}$. At the corresponding laser detunings, the excited macrodimers were detected by their spontaneous ionization rate.  Macrodimers were identified by the unambiguous assignment of the spectroscopic signals to theoretically predicted binding potentials.  Figure adapted from~\cite{Sassmannshausen2016}. The same group also found binding potentials at different principal quantum numbers $n$~\cite{Sassmannshausen20162,Sassmannshausen2015}.}
\end{figure}

\subsection{Identification of binding potentials}

The second observation was reported in 2016, also for $^{133}$Cs~\cite{Sassmannshausen2016}. 
This time, pair states energetically close to the asymptotic state $|nP_{3/2},(n+1)S_{1/2}\rangle$ with $n \approx 45$ were excited in a sequential process, see Fig.~\ref{fig:Deiglmayr}.
A first ``seed" pulse excited atoms in an optical dipole trap from the ground state $\ket{6S_{1/2}}$ into $|(n+1)S_{1/2}\rangle$ on a two-photon transition.
A second pulse drove the single-photon UV transition $\ket{6S_{1/2}} \rightarrow |nP_{3/2}\rangle$ at a detuning $\Delta/(2\pi) = U$ that compensates for the energy shift $U$ between the asymptotic state and the binding potential.
The presence of a seed atom at the right distance then ``facilitated" the excitation of the macrodimer state.
The presence of Rydberg atoms was again verified by detecting the ions after ionization.

Macrodimers were distinguished from single-atom Rydberg excitations because they were found to spontaneously ionize, without applying PFI.
The underlying process might have been auto-ionization~\cite{hahn_ionization_2000,robicheaux_ionization_2005}, possibly triggered by non-adiabatic transitions from the macrodimer state to attractive pair potentials~\cite{Ionization_Gallagher,Amthor_mechanical_ionization}, or by the presence of other nearby Rydberg atoms~\cite{Thesis_Sassmanshausen,Spontaneous_ion_Gallagh,ionization_weidem_2013}.
The detunings $\Delta$ at which spontaneous ionization was observed agreed with the calculated energies $U$ of the minima and maxima of the pair potentials~\cite{Deiglmayr2016}. 
Further studies discussed different PA schemes and compared signatures between interacting Rydberg pair states and Rydberg--ground-state molecules which were observed under similar conditions~\cite{Sassmannshausen20162,Sassmannshausen2015,sasmannshausen_exotic_2016,Thesis_Sassmanshausen}.
\begin{figure}
  \centering
  \includegraphics[width=1.0\columnwidth]{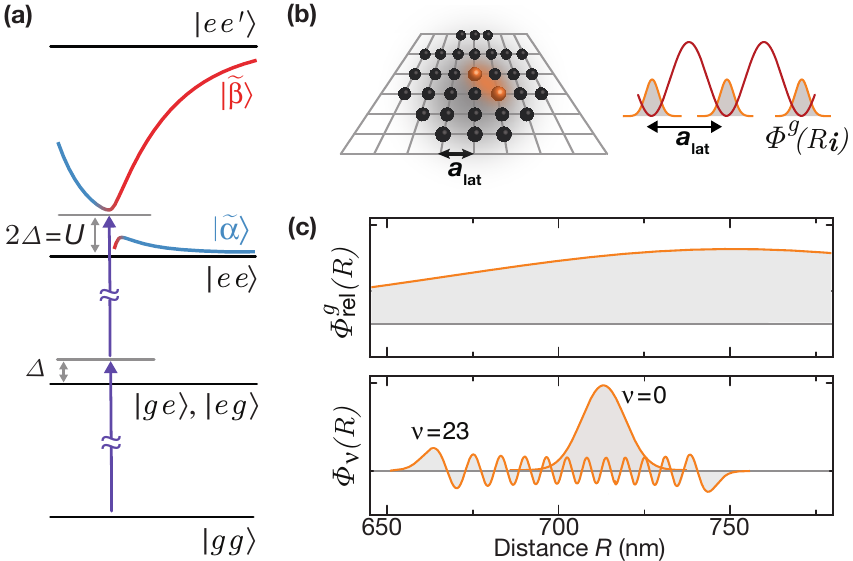}%
  \caption{\label{fig:exc_schem_lat}
  \textbf{Two-photon photoassociation~(PA) with enhanced motional overlap.}  (\textbf{a}) For the rest of the article, ground state atom pairs are photoassociated in a two-photon process from $\ket{gg}$ via intermediate states $\ket{ge}$ and $\ket{eg}$ detuned by a laser detuning $\Delta$ into the doubly-excited macrodimer states. (\textbf{b}) The atoms (black) are arranged in a two-dimensional square array with lattice constant $a_{\mathrm{lat}} = 532\,$nm. This enhances the PA rates of macrodimers whose bond length is close to the lattice diagonal direction (orange). The atoms are prepared in the motional ground state $\Phi^g(R_\mathbfit{i})$ of the traps of the array. (\textbf{c}) The initial relative wave function $\Phi^g_\mathrm{rel}(R)$ is typically significantly broader than the narrow vibrational mode of the macrodimers $\Phi_\nu(R)$. Figure adapted from~\cite{Hollerith_2019}. }
\end{figure}
\subsection{Vibrationally resolved spectroscopy}\label{sec:vibresspec}
\begin{figure*}
  \centering
  \includegraphics[width=1.0\textwidth]{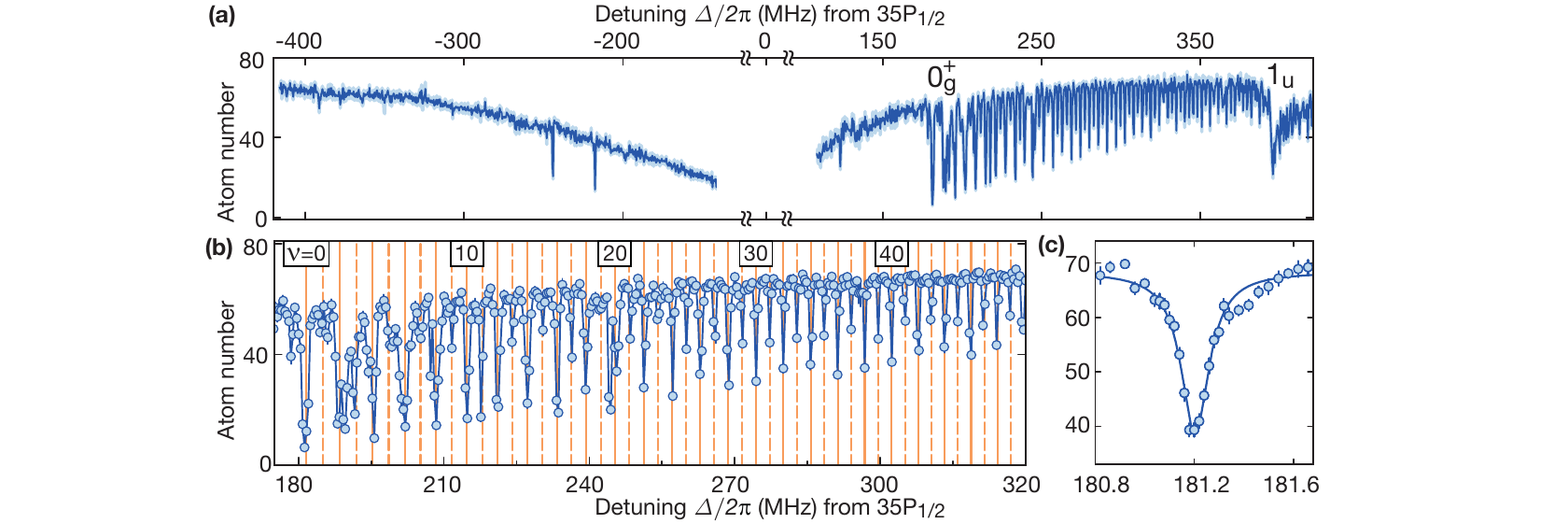}%
  \caption{\label{fig:vib_spec}
  \textbf{Vibrationally resolved macrodimer spectroscopy in an atomic array.}  \textbf{(a)} At negative detunings, the single-photon $35P_{1/2}$ Rydberg resonance is interaction broadened due to the presence of attractive van der Waals potentials. At positive detunings $\Delta/(2\pi) > 180$\,MHz, one observes a series of narrow $0^+_g$ macrodimer resonances excited in a two-photon process from ground state atom pairs. For $\Delta/(2\pi) > 370$\,MHz, another binding potential becomes two-photon resonant. \textbf{(b)} The observed individual lines correspond to spectroscopically-resolved vibrational modes $\nu$. The observations agree with the calculated line positions for even (solid) and odd (dashed) modes $\nu$. \textbf{(c)} The observed resonance profile of the lowest vibrational mode has a full width at half maximum of $139 \pm 5$\,kHz. Figure adapted from~\cite{Hollerith_2019}.}
\end{figure*}
A detailed spectroscopic study of the quantized vibrational states in 2019 for $^{87}$Rb provided compelling evidence of vibrationally bound macrodimer modes~\cite{Hollerith_2019}.
Similar to some of the first studies~\cite{Macrodimers_1,Overstreet2009,Sassmannshausen2015}, macrodimers were excited in a two-photon transition $\ket{gg}\rightarrow \ket{\Psi^\nu_{\textrm{Mol}}}$ in the pair-state basis using UV light, see Fig.~\ref{fig:exc_schem_lat}. 
The macrodimer states excited by a single-frequency narrow-linewidth UV laser become two-photon resonant at detunings $\Delta_\nu/(2\pi) = U_\nu/2$, half as large as the interaction energy $U_\nu$ of the macrodimer states relative to the asymptotic state.
The initially prepared ground state atoms were arranged in a two-dimensional optical square array where each site was typically occupied by one atom. 
The diagonal distance in the array approximately coincided with the minimum of the binding potential. 
The enhanced motional-state overlap $f_\nu$ increased the macrodimer signal while reducing signatures related to Rydberg interactions at other distances.
Such a study was possible because the large macrodimer bond length $R_\nu$ was comparable to the wavelength of the laser light used to create the array.
Contrary to the ion signal used in previous experiments, excited Rydberg states were detected in a quantum gas microscope with single-atom sensitivity and microscopic resolution~\cite{Bakr2009,Sherson2010}. 
Excited macrodimers were observed as atom loss because they are repelled by the light field creating the array and because of the kinetic energy released in their decay~\cite{Thesis_Hollerith}.

The vibrational spectra were observed in the spectroscopic region between the two single-photon resonances from the ground state $\ket{g} = \ket{F=2,m_F = 0}$ into the two fine-structure states $\ket{35P_{1/2}}$ and $\ket{35P_{3/2}}$.
One of the observed spectra is shown in Fig.~\ref{fig:vib_spec}.
The excitation rates were strongest for the lowest vibrational mode $\nu=0$ and decreased for higher even $\nu$.
Coupling rates into odd vibrational states were typically weaker and slightly increase with $\nu$.
This is expected from the Franck-Condon factor $f_\nu$. 
For ground state atoms occupying motional ground states $\Phi^g(R_{\mathbfit{i}})$ of sites $\mathbfit{i}$ of the array, it can be estimated via 
\begin{equation}\label{eq:FC_fac}
f_\nu \approx \int_R \Phi^\star_\nu(R)\Phi^g_\mathrm{rel}(R)dR,
\end{equation}
with $\Phi^g_\mathrm{rel}(R)$ the relative wave function of two atoms after separating the center of mass motion.
The large bond length $R_\nu$ and the small width of the vibrational mode justifies a one-dimensional treatment. 
 
 \subsection{Precision test of Rydberg interactions}
Vibrational spectroscopy of macrodimer modes is the so far most precise experimental test of Rydberg interactions because it provides sharp spectroscopic signatures and probes the non-perturbative regime at short distances.
In the presented studies, the spectroscopic resolution was limited to a few hundred kilohertz~\cite{Thesis_Hollerith}.
For alkali atoms, where quantum defects are well-known and calculations less challenging, the vibrational energies can be calculated at a similarly high precision~\cite{Weber2017,Sibalic2017}, see also section~\ref{sec:theoretical_Des}. 
This requires expanding Eq.~\ref{eq:multipole_expansion} in several thousands of basis states while also exploiting the symmetry of the molecular state.
Furthermore, higher orders in the multipole expansion must be included. 
Approaching the experimental precision in the calculations required to consider at least terms up to octupole-octupole interactions and other terms scaling as $\hat{H}^{(6)}_{\textrm{int}}(\mathbf{R}) \propto R^{-6}$.
The high spectroscopic precision will be further illustrated in the next section where small perturbations in the potentials are discussed.

\section{Non-adiabatic motional couplings}
The Born-Oppenheimer approximation is one of the cornerstones of molecular physics.
However, non-adiabatic transitions between different Born-Oppenheimer potentials, where the approximation breaks down, also play an important role in nature.
Such non-radiative decay channels between different potential surfaces were observed in photochemistry~\cite{photochemistry_conint,blanchet_discerning_1999} and possibly contribute to photobiological processes such as photosynthesis~\cite{photosynth_conint2,photosynthesis_ConInt}.
They also occur in macrodimers~\cite{kiffner_synthetic_so_c_MD}, where they can be studied with a high level of control.

\subsection{Higher-order multipole terms}
\begin{figure}
  \centering
  \includegraphics[width=1.0\columnwidth]{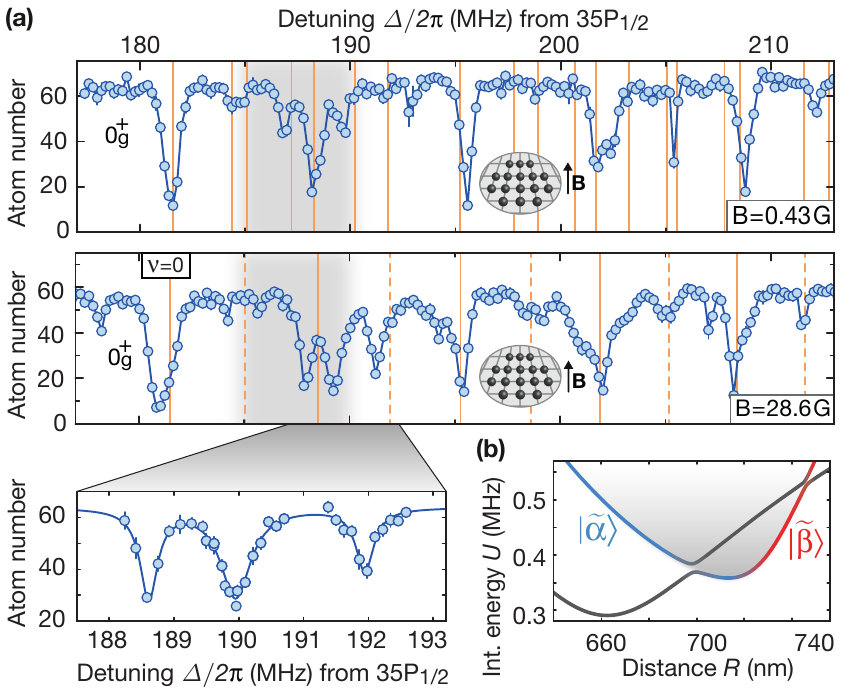}%
  \caption{\label{fig:vibr_cp}
  \textbf{Breakdown of the Born-Oppenheimer approximation.}  \textbf{(a)} Further spectroscopies at different magnetic fields of the vibrational series presented in Fig.~\ref{fig:vib_spec} at low vibrational quantum numbers reveals a deviation from the coarse structure as well as an additional unresolved substructure. For a negligible field amplitude $B=0.43\,$G, most of the observed lines are theoretically predicted (orange lines). For $B = 28.6\,$G where calculations were too challenging, the calculated eigenenergies without accounting for the additional gap are shown. \textbf{(b)} These observations originate from the coupling between two sets of vibrational modes hosted by two crossing pair potentials. Figure adapted from~\cite{Hollerith_2019}.}
\end{figure}

This is illustrated by performing additional precision scans of the lower vibrational modes of the potential discussed in Fig.~\ref{fig:vib_spec}, see Fig.~\ref{fig:vibr_cp}.
Instead of the initially expected regular harmonic oscillator spectrum, one finds a set of broadened resonances at irregular spacings and further unresolved substructure~\cite{Hollerith_2019}.
A closer look into the calculated pair potential reveals an additional $0^+_g$ binding potential asymptotically connected to the non-interacting pair state $\ket{32D_{3/2},37P_{3/2}}$ which crosses the pair potential in the relevant frequency region. 
The crossing induces an additional gap that is energetically comparable to the vibrational energy splitting in the binding potential.

As a consequence, the interatomic motion cannot be restricted to a single Born-Oppenheimer potential.
However, the vibronic eigenstates can still be expanded in a Born-Oppenheimer expansion~\cite{domcke_conical_2004,Pacher_gauge}
\begin{equation}\label{eq:BO_exp}
|\Psi_{\textrm{Mol}}\rangle = \sum_j \bar \Phi_j(R)|\bar \Psi^j_{\textrm{el}}(R) \rangle.
\end{equation}
They are now a superposition of different pair potentials $\ket{\bar \Psi^j_{\textrm{el}}(R)}$ with spatially dependent amplitudes $\bar \Phi_j(R)$.
In the limit of a single potential where the Born-Oppenheimer approximation holds, $\bar \Phi_j(R)$ are the motional states hosted by this potential.

\begin{figure}
  \centering
  \includegraphics[width=1.0\columnwidth]{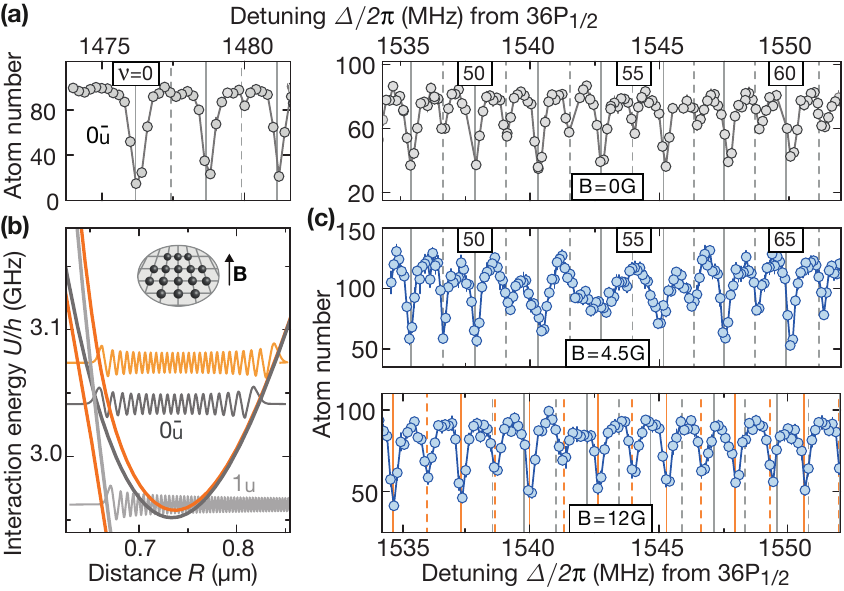}%
  \caption{\label{fig:ct_pred}
  \textbf{Potential engineering and controlled predissociation.} \textbf{(a)} At zero field, the observed spectrum agrees with the calculated vibrational states in the $0^-_u$ potential~(gray lines). \textbf{(b)} A finite magnetic field $\mathbf{B}$ pointing out of the atomic plane is perpendicular to all molecular orientations. 
  This induces a coupling between the $0^-_u$ binding potential~(dark gray) and the repulsive $1_u$ potential~(light gray) that is proportional to $B$. The size of the energy gap can be calculated using degenerate perturbation theory~(orange). \textbf{(c)} For $B = 4.5\,$G, the coupling to the $1_u$ potential broadens some of the vibrational lines. For even stronger $B = 12\,$G, the broadening vanishes. Now, the observed larger vibrational spacing (orange) indicates an adiabatic motion in the combined potential where both pair potentials $0^-_u$ and $1_u$ are mixed. Figure adapted from~\cite{Hollerith_2021}. }
\end{figure}
\begin{figure*}
  \centering
  \includegraphics[width=1.0\textwidth]{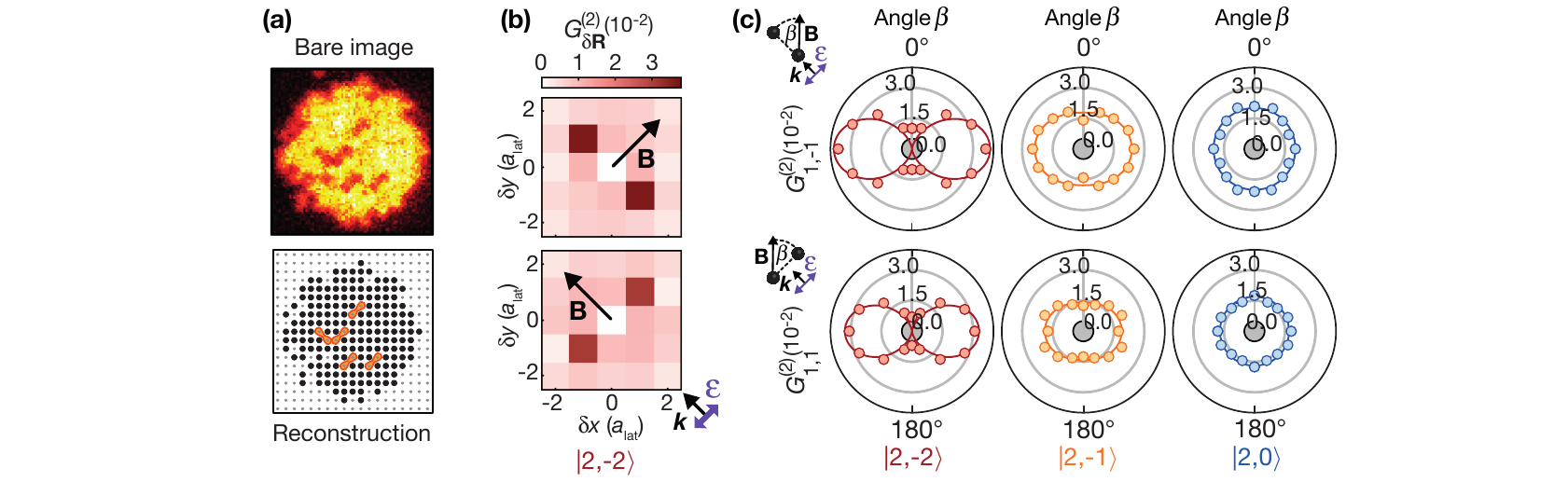}%
  \caption{\label{fig:tomo_1}
  \textbf{Photoassociation in the molecular frame of reference.}  \textbf{(a)} An exemplary image of the two-dimensional atomic array after a photoassociation (PA) pulse. PA can be microscopically resolved by observing correlated atom loss at the lattice diagonal distance, close to the macrodimer bond length. \textbf{(b)} Two correlation signals $G^{(2)}_{\delta\mathbf{R}}$ recorded after exciting ground state atoms $\ket{F=2,m_F=-2}$ into the lowest vibrational resonance of a binding potential observed blue detuned from the $36P_{1/2}$ resonance for two magnetic field orientations. As expected, $G^{(2)}_{\delta\mathbf{R}}$ peaks at the distances $(1,\pm 1) a_{\mathrm{lat}}$. Furthermore, PA rates of molecules oriented parallel to an applied magnetic field $\mathbf{B}$ vanish. \textbf{(c)} The dependence of $G^{(2)}_{1,-1}$ with $\mathbf{R}\perp\mathbfit{\varepsilon}$ and $G^{(2)}_{1,+1}$ with $\mathbf{R}\parallel\mathbfit{\varepsilon}$ for initial states $\ket{2,-2}$~(red), $\ket{2,-1}$~(orange), and $\ket{2,0}$~(blue) on the angle $\beta$ between $\mathbf{B}$ and $\mathbf{R}$ is characteristic for $0^-_u$ potentials. The solid lines represent the theoretical expectation, the overall signal strength was left as a fit parameter. Figure adapted from~\cite{Hollerith_2019} and~\cite{Hollerith_2021}.}
\end{figure*}

Because of the large density of Rydberg pair states, such crossings between pair potentials occur frequently. 
The coupling between crossing pair potentials depends on the involved multipole term in Eq.~\ref{eq:multipole_expansion} and can generally also depend on external fields and their orientation relative to the interatomic axis.
Dipole-dipole interactions $\hat{H}^{(3)}_{\mathrm{int}}(\mathbf{R})$ typically induce large gaps where the motion follows the avoided crossings at the gap adiabatically.
In the case discussed here, the coupling is induced by the dipole-quadrupole interaction term $\hat{H}^{(4)}_{\mathrm{int}}(\mathbf{R})$.
For smaller gaps induced by even high-order multipole terms, the vibrating molecule typically does not recognize the gap and follows the crossing diabatically. 

For small magnetic fields, the observed line structure can be calculated by accounting for the non-adiabatic motional couplings. 
To first-order, both $0^+_g$ potentials and the gap are insensitive to $B$.
At higher fields, the vibrational energies experience a small second-order shift. 
This affects the vibronic structure in the combined potential because it depends on the exact location of the crossing relative to the potential minimum.

\subsection{Magnetic field induced predissociation}\label{magn_predis}
Experimental signatures related to a breakdown of the Born-Oppenheimer approximation depend on the binding or non-binding character of the potentials involved.
So-called predissociation emerges when bound vibrational modes are coupled to a continuum of unbound states that reduce their lifetime below the radiative lifetime of both contributing Rydberg states~\cite{Predissociation_Chemical_Rev}.

This paragraph discusses macrodimer predissociation controlled by magnetic fields~\cite{Hollerith_2021}.
A vibrational spectrum of $0^-_u$ macrodimers in the vicinity of a crossing repulsive $1_u$ potential is shown in Fig.~\ref{fig:ct_pred}.
In the absence of external field components perpendicular to the interatomic axis $R$, both potentials are uncoupled because the angular momentum projection $\Omega$ is conserved.
However, if a field breaks the cylindrical symmetry of the molecule, a coupling becomes possible. 
In contrast to the discussion of Fig.~\ref{fig:vibr_cp}, the gap now strongly depends on the field amplitude.
The coupling to the motional continuum and the reduced lifetime of the macrodimer states leads to a broadening of some of the vibrational levels.
Similar effects have been observed for more deeply bound molecules~\cite{vigue_j_natural_1981,vigue_j._natural_1981_2,magnetic_prediss}.
At very high magnetic fields, the broadening disappears.
Now, the vibrational motion follows the avoided crossing adiabatically.

\section{Electronic structure tomography}\label{sec:tomo}
The vibrational spectra not only contain information about the binding energies but also about the electronic quantum numbers $|\Omega|^\pm_{g/u}$.
Dependent on the angular momentum projection of the hyperfine ground state, some molecular potentials can be coupled while others remain uncoupled~\cite{Hollerith_2019,Hollerith_2021}.
An even more detailed study of the electronic structure becomes possible using the spatial arrangement introduced in Fig.~\ref{fig:exc_schem_lat}.
The relative orientation of all atom pairs is well-defined.
The wave function of the relative orientation can be expanded into spherical harmonics $\ket{\psi_{\mathrm{or}}} = \sum_{\ell m} c_{\ell m}\ket{\ell m}$, with coefficients $c_{\ell m}$ that depend on the individual traps in the array.
As discussed in section~\ref{properties}, the rotational states are effectively degenerate.
Because the couplings to rotational states from the ground state are proportional to the amplitudes $c_{\ell m}$, the excited macrodimers will be in a superposition of rotational states that preserves the orientation.
The interatomic axis which serves as a quantization axis for the electronic structure is therefore aligned in the laboratory frame.

Furthermore, because of their large distance, the photoassociated atoms can be microscopically resolved~\cite{Hollerith_2019,Hollerith_2021}. 
This has been realized using the site-resolved fluorescence imaging of the quantum gas microscope mentioned in section~\ref{sec:vibresspec}.
An exemplary image and the reconstructed atom occupations are shown in Fig.~\ref{fig:tomo_1}\,(a).
The microscopically resolved excitation rates can be quantified by evaluating correlations
\begin{equation}\label{eq:g2}
G^{(2)}_{\delta \mathbf{R}} = \left(\langle \hat{h}_{\mathbf{R}^\prime + \delta \mathbf{R}}\hat{h}_{\mathbf{R}^\prime} \rangle - \langle \hat{h}_{\mathbf{R}^\prime + \delta \mathbf{R}}\rangle\langle\hat{h}_{\mathbf{R}^\prime} \rangle\right)_{\mathbf{R}^\prime}
\end{equation}
between empty sites at distance $\delta \mathbf{R}$ in the reconstructed images.
Here, $\left(\,.\,\right)_{\mathbf{R}^\prime}$ denotes averaging over all sites $\mathbf{R}^\prime$ of the array and $\langle \, . \, \rangle$ averaging over experimental realizations.
The projector $\hat{h}_{\mathbf{R}^\prime}$ provides $1$~($0$) for an empty~(occupied) site at position $\mathbf{R}^\prime$. 

\subsection{Identifying molecular symmetries}
This combination of molecular alignment and microscopic access enables PA studies, where the molecular orientation relative to external fields, the light polarization, and the initial atomic state are fully controlled. 
This paragraph discusses the dependence of PA on the angle $\beta$ between an applied field $\mathbf{B}$ and the interatomic axis $\mathbf{R}$.
If the initially unbound atoms have a well-defined angular momentum projection relative to $\mathbf{B}$, the Clebsch-Gordan coefficients contributing to the optical coupling provide characteristic dependencies for different molecular states $|\Omega|^\pm_{g/u}$.
This was studied at a $0^-_u$ potential observed blue-detuned from the UV transition $\ket{g}\rightarrow\ket{36P_{1/2}}$ in $^{87}$Rb~\cite{Hollerith_2021}. 
The laser was resonant with the lowest vibrational mode.
After photoassociating a few molecules from the atoms prepared in the hyperfine ground state $\ket{F=2,m_F=-2}$, the remaining atoms in the array were imaged.
Two exemplary correlations for two orthogonal orientations of the magnetic field $\mathbf{B}$ are shown in Fig.~\ref{fig:tomo_1}\,(b).
In both cases, the molecular signal reveals an alignment perpendicular to $\mathbf{B}$.
A more rigorous measurement of the PA rate for a varying angle $\beta$ shows different functional dependencies for initial states $|F,m_F\rangle$.
The characteristic curves follow the theoretical expectations for $0^-_u$ potentials -- assuming the excitation of $^{87}\mathrm{Rb}$ pairs using the scheme introduced in Fig.~\ref{fig:exc_schem_lat}\,(\textbf{a}).
For other potentials $\Omega^{\pm}_{g/u}$, other characteristic curves are expected.

The observations presented in Fig.~\ref{fig:tomo_1} also indicate sligthly stronger excitation rates for light polarizations $\mathbfit{\varepsilon}\perp \mathbf{R}$ compared to $\mathbfit{\varepsilon}\parallel \mathbf{R}$.
This is in agreement with the calculated two-photon excitation rates accounting for the contributing pair states in Eq.~\ref{eq:pair_pot_exp} for this specific $0^-_u$ potential.

\subsection{Response to magnetic fields}
\begin{figure}
  \centering
  \includegraphics[width=1.0\columnwidth]{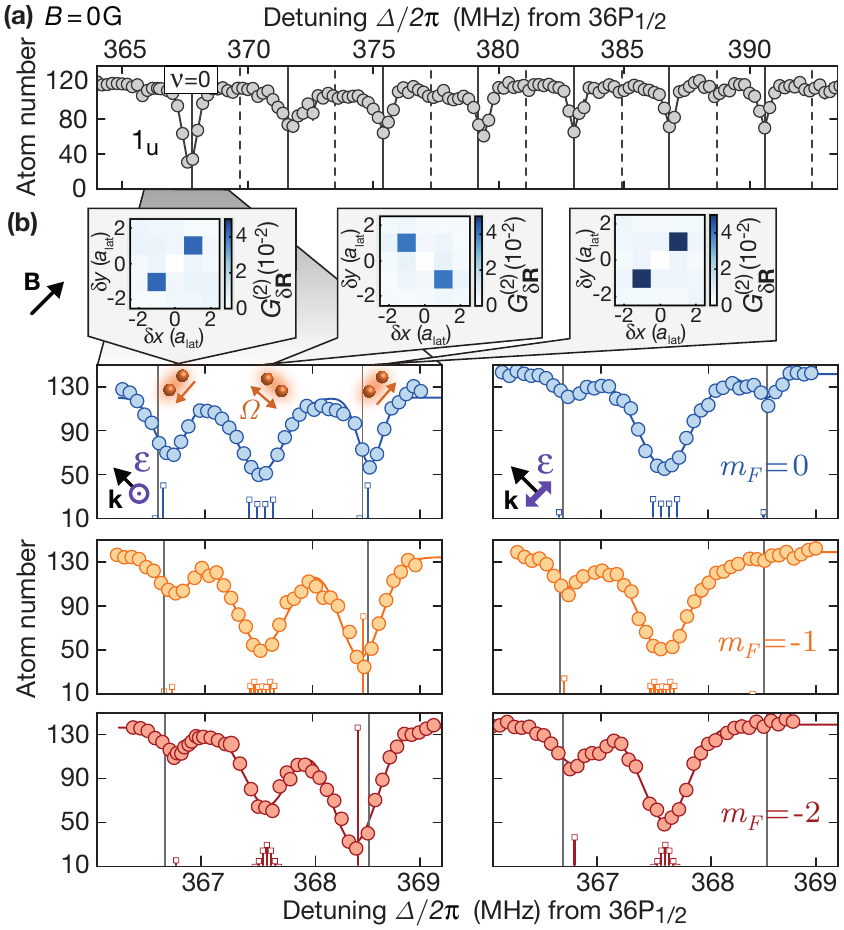}%
  \caption{\label{fig:tomo_2}
  \textbf{Macrodimer Zeeman and hyperfine interaction.}  \textbf{(a)} At zero field, the vibrational spectrum of the $1_u$ binding potential agrees with the calculation (gray lines). \textbf{(b)}
High-resolution spectroscopy of the lowest vibrational line for a magnetic field $B=1.0\,$G along one of the diagonal directions reveals a splitting into three lines.  The recorded correlation signals reveal an molecular alignment $\mathbf{R}\parallel\mathbf{B}$ at the outer resonances and $\mathbf{R}\perp\mathbf{B}$ at the central resonance. 
Scans for different initial states $\ket{F = 2,m_F = 0}$~(blue), $\ket{2,-1}$~(orange) and $\ket{2,-2}$~(red) and light polarizations $\mathbfit{\varepsilon}\parallel \mathbf{B}$ and $\mathbfit{\varepsilon}\perp \mathbf{B}$ show a small but significant deviation from the calculated splitting accounting only for the electronic macrodimer structure (vertical gray lines). After including also the hyperfine interactions of the pair potentials, the observations can be explained (colored bars). The bar height indicates the calculated relative strength of the lines. Figure adapted from~\cite{Hollerith_2021}. }
\end{figure}
Microscopic detection via correlation measurements furthermore enables studying macrodimers by their response to external fields. 
For atoms, different angular momentum projections typically split in the presence of an external field, which then acts as an external quantization axis.
For molecules, however, the interatomic binding provides an internal quantization axis. 
The response of the electronic wave function to external fields now depends on its orientation relative to the field.
This has been studied for a $1_u$ molecular potential which was again located blue-detuned from the $\ket{g}\rightarrow\ket{36P_{1/2}}$ resonance~\cite{Hollerith_2021}.
In contrast to molecular potentials with $\Omega = 0$, one finds two pair potentials $\Omega = \pm 1$ that are degenerate at zero field. 
Applying a magnetic field along one diagonal direction of the array shows a splitting of the vibrational resonances into three lines. 
The recorded correlation signals reveal that molecules excited at the two outer resonances are oriented parallel to $\mathbf{B}$ while molecules at the central unshifted line are oriented perpendicular.
This agrees with a calculation in first-order perturbation theory which predicts an energy shift between $\Omega = \pm 1$ proportional to the field component parallel to $\mathbf{R}$.

This observation again relies on the alignment of the ground state atoms and the excited macrodimers, provided by the lattice: 
For randomly oriented macrodimers, one would observe a broadening instead of a line-splitting.
Both cases differ from the Zeeman splitting of conventional molecules occupying a low and well-defined rotational state. 
Here, one again observes a set of quantized Zeeman lines because of the limited amount of projections of the contributing coupled angular momenta on the magnetic field axis~\cite{Molecular_Zeeman2,Molecular_Zeeman,Shade_molecular_Zeeman,Zeeman_shift_molecules}.

 Measuring the line splittings for $\mathbf{R}\parallel\mathbf{B}$ from the non-shifted central reference line with $\mathbf{R}\perp\mathbf{B}$ also enables a more quantitative analysis.
The calculated splitting obtained in first-order perturbation theory using the electronic wave function Eq.~\ref{eq:pair_pot_exp} provides reasonable agreement with the experiments.
The remaining deviation depends on the initial hyperfine state $\ket{F=2,m_F}$ as well as the light polarization, which is a striking indication of hyperfine interactions in the Rydberg manifold.
Extending the calculation by the hyperfine interaction of the Rydberg pair states contributing to the macrodimer state explained the observations, showing once more the relevance of macrodimer spectroscopy for benchmarking Rydberg pair potentials. 
As expected, studies at lower principal quantum numbers $n$ revealed an even stronger contribution of the hyperfine interaction~\cite{Hollerith_2021,Thesis_Hollerith}. 

The line strengths of the outer resonances with $\mathbf{R}\parallel\mathbf{B}$  can be explained by the Clebsch-Gordan coefficients contributing to the PA, which are larger for a light polarization perpendicular to the interatomic axis $\mathbfit{\varepsilon}\perp\mathbf{R}$ compared to $\mathbfit{\varepsilon}\parallel\mathbf{R}$.
The coupling of angular momentum states also correctly predicts the asymmetry between both outer lines for ground states $m_F\neq 0$.

\subsection{Spatially varying electronic structure}
\begin{figure}
  \centering
  \includegraphics[width=1.0\columnwidth]{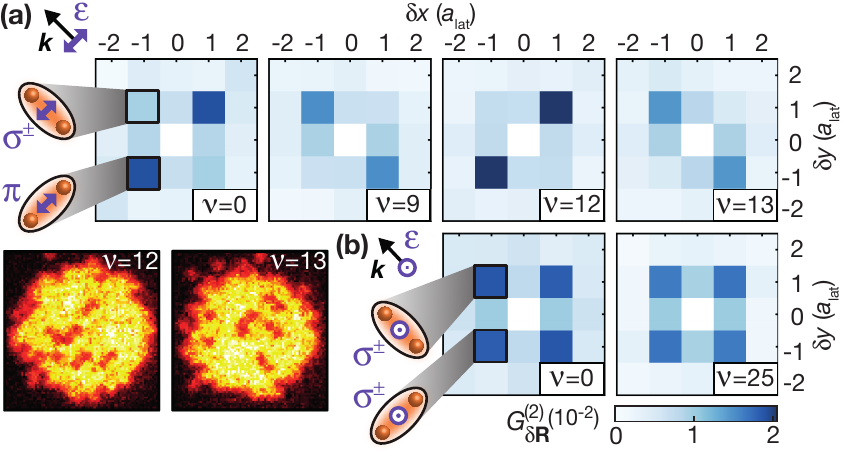}%
  \caption{\label{fig:tomo_3}
  \textbf{Molecular alignment via vibrational states.} (\textbf{a}) The correlation signal at the different vibratonal resonances discussed in Fig.~\ref{fig:vib_spec} shows stronger PA rates for $\mathbf{R}\parallel \mathbfit{\varepsilon}$~($\mathbf{R}\perp\mathbfit{\varepsilon}$) for even~(odd) $\nu$. This can be explained by accounting for the parametric $R-$dependence of the electronic structure. The orientation of the excited molecules can also be seen in the individual images. (\textbf{b}) Rotating the light polarization $\mathbfit{\varepsilon}$ out of the atomic plane, the excitation rate becomes symmetric along both directions. Figure adapted from~\cite{Hollerith_2019}. }
\end{figure}
\begin{figure}
  \centering
  \includegraphics[width=1.0\columnwidth]{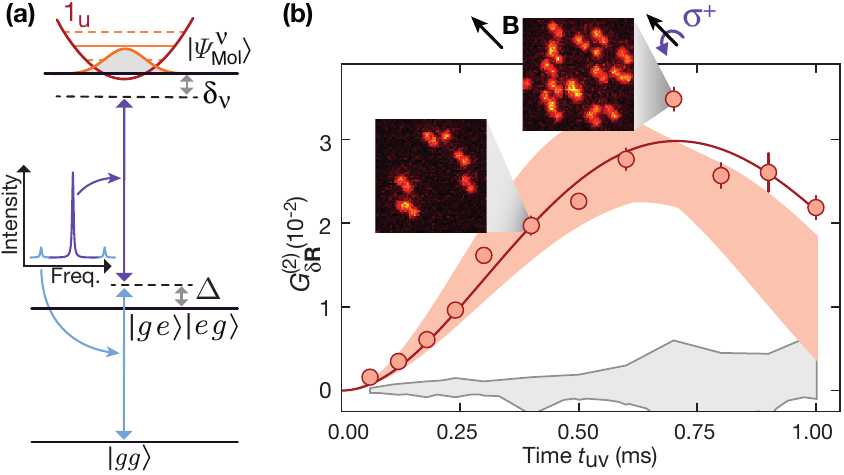}%
  \caption{\label{fig:tomo_2c1}
  \textbf{Distance-selective interactions using macrodimers.}  (\textbf{a}) Macrodimer excitation using one modulated sideband photon (light blue) and one carrier photon (purple) provides tunable intermediate state detunings $\Delta$ to singly-excited intermediate states $\ket{ge}$ and $\ket{eg}$ with $\ket{g} = \ket{F=2,mF=-2}$ and $\ket{e} = \ket{36P_{1/2}, m_J = +1/2}$. For the chosen initial state $\ket{g}$ and a $\sigma^{+}$ polarized excitation field, the coupling rates into the $1_u$ macrodimers presented in Fig.~\ref{fig:tomo_2} reach a maximum for orientations $\mathbf{R_\parallel}$ parallel to $\mathbf{B}$, while $\mathbf{R_\perp}$ is suppressed. Strong dressed interactions are realized for small intermediate state detunings $\Delta$ and two-photon detunings $\delta_\nu$ to macrodimer states $\ket{\Psi^\nu_{\textrm{Mol}}}$. (\textbf{b}) Time-dependent correlations for $\delta\mathbf{R} = \mathbf{R}_\parallel$ observed in a Ramsey sequence (red data points). For increasing dressing time $t_{\mathrm{uv}}$, the interactions induce correlated spin flips between spin pairs at the lattice diagonal distance which is close to the binding potential minimum. The solid red line represents a fit from where the experimental spin coupling was extracted, the red shaded area indicates the theoretical expectation. The gray area shows the background signal between spin pairs that are not coupled to molecular states. Two exemplary images from the quantum gas microscope are included. Figure adapted from~\cite{hollerith2021realizing}.}
\end{figure}

As for any molecule, the electronic structure of macrodimers depends on the interatomic distance $R$ because the admixtures $c_{ij}(R)$ of non-interacting pair states in Eq.~\ref{eq:pair_pot_exp} mediated by the interaction Hamiltonian depend on $R$.
Within the Born-Oppenheimer framework, this means that the electronic wave function depends parametrically on $R$~\cite{domcke_conical_2004}. 
For the potential introduced in Fig.~\ref{fig:schematic}, the electronic wave function is periodically transferred from $\ket{\widetilde\alpha}$ to $\ket{\widetilde\beta}$ during the vibrational motion.
\begin{figure*}
  \centering
  \includegraphics[width=0.75\textwidth]{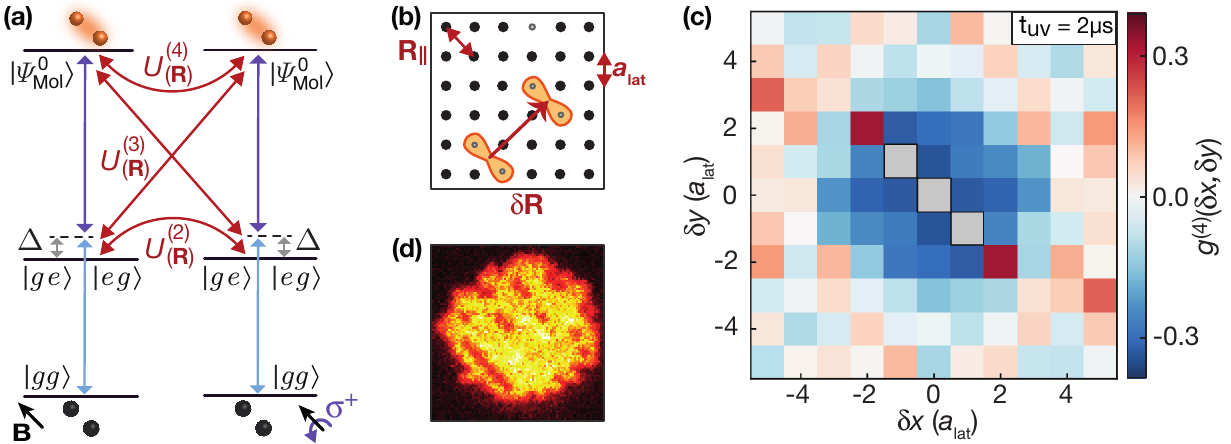}%
  \caption{\label{fig:mac_block}
  \textbf{Evidence of a macrodimer blockade.} (\textbf{a}) Pairs of ground state atoms are resonantly excited to macrodimer states $\ket{\Psi^0_{\textrm{Mol}}}$ with $\nu = 0$. The initial state, the magnetic field orientation, the light polarization, and the binding potential were identical to Fig.~\ref{fig:tomo_2c1}. The intermediate state detuning was set to $\Delta/(2\pi) = 3.6(1)\,$MHz using again a sideband modulated to the UV laser. Additional Rydberg interactions $U^{(n)}_{(\mathbf{R})}$, with $n$ the number of participating Rydberg states, are expected to suppress the excitation of both pairs at close distances. (\textbf{b}) Two excited macrodimers create two pairs of empty sites in the initially unity-filled array. Their relative distance distribution can be probed via the four-hole correlator $g^{(4)}_{\delta\mathbf{R}}$, with $\delta\mathbf{R}$ the distance between both hole pairs. (\textbf{c}) As expected, the recorded $g^{(4)}_{\delta\mathbf{R}}$ signal shows a reduction of the excitation rate at close distances. The signal was excluded at distances where both pairs share one or two array sites~(gray). Surprisingly, the signal also shows correlated events of four empty sites aligned to the lattice diagonal direction $\mathbf{R}_\parallel$ where the macrodimers are excited. (\textbf{d}) Individual images also indicate the presence of such a loss process. The underlying mechanism is under current investigation. }
\end{figure*}

Because the previous discussion focused on the lowest and spatially narrow vibrational levels $\nu = 0$ of different binding potentials, this effect has been neglected so far.
More generally, the distance-dependent electronic state decomposition has implications for the photoassication process.
Comparing the correlation signals $G^{(2)}_{\delta\mathbf{R}}$ at different vibrational modes $\nu$ of the spectrum presented in Fig.~\ref{fig:vib_spec} provides an illustrative example~\cite{Hollerith_2019}.
The polarization $\mathbfit{\varepsilon}$ of the excitation light was pointing along one lattice diagonal direction of the array.
The signal shows that the alignment of the excited molecules was predominantly $\mathbf{R}\parallel\mathbfit{\varepsilon}$ at even modes $\nu$, while it was mainly $\mathbf{R}\perp\mathbfit{\varepsilon}$ at odd $\nu$.
Because the electronic decomposition barely changes for different modes $\Phi_\nu(R)$, this dependence on the vibrational state contradicts the picture of a single Franck-Condon integral contributing as an overall prefactor to the electronic coupling. 

Instead, the coupling rate $\widetilde\Omega_{\nu} = \widetilde\Omega_{\widetilde\alpha}f_{\widetilde\alpha} +  \widetilde\Omega_{\widetilde\beta}f_{\widetilde\beta}$ from the ground state is a sum of two mixed contributions.
The electronic coupling terms $\widetilde\Omega_{\widetilde\alpha}$ and $\widetilde\Omega_{\widetilde\beta}$ mediated by the light field are weighted by independent spatial integrals $f_{\widetilde\alpha} = \int_R \Phi^\star_\nu(R)c_{\widetilde\alpha}(R)\Phi^g_\mathrm{rel}(R)dR$ and $f_{\widetilde\beta}$.
At the sharp avoided crossing, the coefficients $c_{\widetilde\alpha}(R)$ and $c_{\widetilde\beta}(R)$ of the electronic wave function $|\Psi_{\textrm{el}}(R) \rangle = c_{\widetilde\alpha}(R)|\widetilde\alpha \rangle + c_{\widetilde\beta}(R)|\widetilde\beta \rangle$ rapidly change from the left to the right side of the potential, see also Fig.~\ref{fig:exc_schem_lat}.
Here, for a broad motional ground state $\Phi^g_\mathrm{rel}(R)$, the relative sign of $f_{\widetilde\alpha}$ and $f_{\widetilde\beta}$ equals the parity of the vibrational mode $\nu$.
Achieving constructive interference between both terms contributing to the coupling furthermore depends on the relative sign of  $\widetilde\Omega_{\widetilde\alpha}$ and $\widetilde\Omega_{\widetilde\beta}$, which turns out to be identical~(opposite) for $\mathbf{R}\parallel\mathbfit{\varepsilon}$ ($\mathbf{R}\perp\mathbfit{\varepsilon}$).
Rotating the polarization out of the plane, the system becomes symmetric in the plane.
Now, both molecular orientations were excited at identical rates.  
For the measurements described in this paragraph, the magnetic field was pointing out of the plane and was irrelevant for the experiment.
The additional avoided crossing in the same potential discussed in Fig.~\ref{fig:vibr_cp} and the related non-adiabatic motional couplings also had no relevant effect.

\section{A resource for quantum science}

During the last decade, Rydberg interactions triggered a variety of applications in quantum science. 
These include the realization of long-range interacting many-body Hamiltonians~\cite{Ryd_sim_1,symmetry_p_phase_2019,ebadi_quantum_2021,Geier_XYZ,Chew2022}, quantum gates for quantum computation~\cite{Qcomp_Ryd_atoms_lukin2000,Qcomp_Ryd_atoms_lukin2001,Saffman_review2010,Bluvstein2022,Graham2022-at}, as well as strong photon-photon interactions~\cite{Int_phot_1,Int_phot_2,Free_space_QED,thompson_symmetry-protected_2017}. 
Because macrodimers consist of Rydberg atoms, similar phenomena are expected to be present for macrodimers.

One example is Rydberg dressing where long-range interactions between ground state atoms are realized by off-resonantly coupling to Rydberg states using laser light~\cite{Zeiher2016a,Jau2016,Dress_Monika_transv,Bakr_itinerant_hopping,Lea_dressing}.
Similar schemes have recently been realized by off-resonantly coupling to macrodimer binding potentials~\cite{hollerith2021realizing,vanBijnen2015}, see Fig.~\ref{fig:tomo_2c1}.
The admixed interactions inherit the orientation dependent coupling rates discussed in section~\ref{sec:tomo}.
Compared to the excitation scheme presented in Fig.~\ref{fig:exc_schem_lat}, a new scheme using a phase-modulated UV laser provided smaller and tunable intermediate state detunings.
In contrast to conventional dressing schemes where interactions are present over a large distance regime, the interactions are now only present within a narrow distance window where the Franck-Condon overlaps with the vibrational macrodimer modes are non-negligible.
This dependence on the motional state overlap also requires controlling the motional state of the ground state.
In future studies in optical tweezer arrays~\cite{Bluvstein2022} where evaporative cooling techniques cannot be applied, this can be achieved by using Raman sideband cooling~\cite{Lea_dressing}.

Another example is Rydberg blockade~\cite{urban_rydblock_2009,schaus_crystal_2012} where multiple Rydberg excitations within a volume where interaction shifts are larger than the optical coupling rate from the ground state are suppressed. 
This has been tested using similar experimental conditions as in Fig.~\ref{fig:tomo_2c1} where coupling rates reach a maximum for orientations $\mathbf{R_\parallel}=(1,-1)\,a_{\mathrm{lat}}$.
Now, however, the laser was two-photon resonant to the lowest vibrational mode $\nu = 0$ where $\delta_{0} = 0$.
At the chosen parameters, macrodimers were excited faster than the motional timescale on which they leave their position or their expected lifetime $\tau_{\mathrm{mol}} \approx 20\,\si{\micro\second}$~\cite{Hollerith_2019}. 
We therefore recorded $\approx 21\,000$ images after illuminating the initial atomic array for $t_{\mathrm{uv}} = 2\,\si{\micro\second}$.
During this time we excited roughly $1.5$ macrodimers in a region of interest of $11 \times 11$ sites.
The macrodimer-macrodimer blockade signal can be observed by evaluating the connected four-hole correlation function 
\begin{align}
& g^{(4)}_{\delta\mathbf{R}} = \biggl( \biggr< \frac{ (\hat{h}_{\mathbf{R}^\prime}-\langle\hat{h}_{\mathbf{R}^\prime}\rangle )(\hat{h}_{\mathbf{R}^\prime+\mathbf{R}_{\parallel}}-\langle\hat{h}_{\mathbf{R}^\prime+\mathbf{R}_{\parallel}}\rangle ) }{\langle\hat{h}_{\mathbf{R}^\prime}\rangle\langle\hat{h}_{\mathbf{R}^\prime+\mathbf{R}_{\parallel}}\rangle} ~ \times \\
&
\frac{( \hat{h}_{\mathbf{R}^\prime+\delta\mathbf{R}}-\langle\hat{h}_{\mathbf{R}^\prime+\delta\mathbf{R}}\rangle ) (\hat{h}_{\mathbf{R}^\prime+ \mathbf{R}_\parallel+\delta\mathbf{R}}-\langle\hat{h}_{\mathbf{R}^\prime+ \mathbf{R}_\parallel+\delta\mathbf{R}}\rangle )}{ \langle\hat{h}_{\mathbf{R}^\prime+\delta\mathbf{R}}\rangle\langle\hat{h}_{\mathbf{R}^\prime+\mathbf{R}_{\parallel}+\delta\mathbf{R}}\rangle}\biggr>\biggr) _{\mathbf{R}^\prime}, \notag
\end{align}
with conventions being identical as in Eq.~\ref{eq:g2}.
Two of the three distances contributing to $g^{(4)}_{\delta\mathbf{R}}$ were fixed to $\mathbf{R}_\parallel$, only $\delta\mathbf{R}$ was varied.
Using this correlation function, we conclude the presence of Rydberg blockade for macrodimer states with a blockade radius $r_b \approx 3\,a_{\mathrm{lat}}$ at which the simultaneous excitation of more than one macrodimer is suppressed, see Fig.~\ref{fig:mac_block}.
It is comparable to the conventional Rydberg blockade at similar principal quantum numbers.
A quantitative prediction will require to account for three- and four-atom Rydberg interactions affecting both macrodimers.
Furthermore, the intermediate state experiences interactions involving two and three Rydberg states.
At large distances where the additional interactions only weakly shift the energy of both macrodimers, the interaction might be calculated perturbatively.
At short distances, a full diagonalization of all Rydberg interactions will be required.

Future studies of collectively enhanced Rabi oscillations, similar as the ones observed for Rydberg atoms~\cite{Zeiher_superatom}, might be challenging because deexciting macrodimers can populate other motional states than the one initially populated by the electronic ground state.
Solving these additional challenges, in principle, enables the realization of four-qubit gates using similar schemes as realized for Rydberg atoms~\cite{Saffman_review2010,Saffman_gate,Gate_Lavine_2019}. This requires replacing the couplings between two qubits and Rydberg states with couplings between two pairs of qubits and macrodimers.

Macrodimers might also enable entirely new applications.
Their excitation creates atom pairs whose motional relative wave function is more narrow than in typical optical traps.
It also realizes a distance-selective pair loss which might be capable of engineering novel dissipative many-body systems in optical lattices~\cite{Ates2012,Diss_Loss_Rempe_2008,Barbier2021}.
Furthermore, macrodimers can be used to entangle distant nuclear spins~\cite{Hollerith_2021,Hollerith_2019}.
The formalism describing macrodimer excitation~\cite{Optical_coupling_macrodimers,Hollerith_2021} also enables to estimate the significance of steep potentials at close distances for quantum simulations and computations~\cite{magic_distances_block}.

\section{Conclusion and outlook}
Macrodimers are weakly-bound micrometer-sized diatomic molecules consisting of highly-excited atoms~\cite{Boisseau2002,Overstreet2009,Sassmannshausen2016}.
Their main conceptual difference from conventional molecules is the absence of overlapping electron orbitals as well as the small rotational splittings, which remains unresolved over their radiative lifetime.
Their large size and their comparatively simple theoretical description enable microscopically resolved studies of molecules that can be fully described using the language of atomic physics.
Their vibrational spectra provide the so far most stringent tests of Rydberg interactions and even reveal small perturbations -- such as non-adiabatic motional couplings between different Born-Oppenheimer potentials or the hyperfine interaction of the Rydberg pair potentials~\cite{Hollerith_2019,Hollerith_2021}.

Future studies for non-alkali atoms with more complicated interactions might become valuable to verify theoretical models.
Also shaping macrodimer binding potentials by microwave fields~\cite{Petrosyan_MW_dressing,sevincli_microwave_2014} or by placing atom pairs close to conducting surfaces might be possible~\cite{Casimir_Macrodiemrs}.
Furthermore, Rydberg interaction potentials can also be extended to  larger atom numbers.
In addition to the discussed interaction between pairs of macrodimers, also Förster resonances of three~\cite{Three_body_F_1,Three_body_F_2,Three_body_F_3,Cheinet_2020} and four~\cite{Four_body_F_1} atoms as well as Rydberg aggregates~\cite{Rydberg_aggregates_1,Rydberg_aggregates_2,Rydberg_aggregates_3,Urvoy2015,Resonant_dipole_barredo_2015,symmetry_p_phase_2019} have already been observed.
Finally, the existence of macrotrimers~\cite{macrotrimers_1,macrotrimers_2} -- bound states of three Rydberg atoms -- is predicted.

\section*{Acknowledgements}
The authors thank Jun Rui, Johannes Deiglmayr, and James Shaffer for comments on the manuscript, as well as Sebastian Weber, Valentin Walther, Andreas Kruckenhauser, Dan M. Stamper-Kurn, Pascal Weckesser, Kritsana Srakaew, and Roman Bause for discussions.
The authors acknowledge funding by the Max Planck Society (MPG) and from Deutsche Forschungsgemeinschaft (DFG, German Research Foundation) under Germany’s Excellence Strategy – EXC- 2111 – 390814868 and Project No. BL 574/15-1 within SPP 1929 (GiRyd). This project has received funding from the European Union’s Horizon 2020 research and innovation programme under grant agreement No. 817482 (PASQuanS). J.Z. acknowledges support from the BMBF through the program “Quantum technologies - from basic research to market” (Grant No. 13N16265).

\bibliography{Review}

\end{document}